\def\gr{$\gamma$-ray }
\def\grs{$\gamma$-rays }

\documentstyle[12pt,aaspp4]{article}

\begin{document}

\title{Electron acceleration in SNR and diffuse gamma-rays above 1 GeV}
\author{Martin Pohl}
\affil{Danish Space Research Institute, Juliane Maries Vej 30, 2100 Copenhagen \O, Denmark}
\author{Joseph A. Esposito}
\affil{GSFC, Code 662, Greenbelt, MD 20771, USA}
\authoremail{mkp@dsri.dk}

\begin{abstract}
The recently observed X-ray synchrotron emission
from four supernova remnants (SNR) has strengthened the evidence
that cosmic ray electrons are accelerated in SNR.
We show, that if this is indeed the case, the local electron
spectrum will be strongly time-dependent, at least above roughly 30 GeV.
The time dependence stems from the Poisson
fluctuations in the number of SNR within a certain volume and 
within a certain time interval. As far as cosmic ray electrons
are concerned, the Galaxy looks like actively bubbling
swiss cheese rather than a steady, homogeneously filled system.

Our finding has important consequences for studies of the 
Galactic diffuse gamma-ray emission, for which a strong excess
over model predictions above 1 GeV has been reported recently.
While these models were relying on an electron injection spectrum
with index 2.4 -- chosen to fit the local electron flux up to
1 TeV -- we show that an electron injection index of around 2.0 would
a) be consistent with the expected Poisson fluctuations in the
locally observable
electron spectrum and b) explain the above mentioned
gamma-ray excess above 1 GeV. An electron injection index around 2
would also correspond to the average radio synchrotron spectrum
of individual SNR. We use a three-dimensional propagation code
to calculate the spectra of electrons throughout the Galaxy and show
that the longitude and latitude distribution of the leptonic gamma-ray
production above 1 GeV is in accord with the respective distributions
for the gamma-ray excess.

We finally point out that our model implies a strong systematic
uncertainty in the determination of the spectrum of the extragalactic
gamma-ray background.

\end{abstract}


\section{Introduction}
As was first observed by OSO-3 (\cite{kraus}), the dominant feature
of the high-energy \gr sky is the intense emission from the Galactic
plane. Later the complete SAS-2 (\cite{ficht75}) and COS-B (\cite{mh80})
data gave evidence for a correlation between the \gr emission
and the spatial structures of the Galaxy. The intensity distribution 
and the spectral form of the emission have led to the consensus that
the diffuse \gr radiation is primarily produced by interactions between
Galactic cosmic ray particles and the interstellar medium, and to a small
extent by unresolved Galactic point sources (\cite{bloe89,str95}).
The EGRET observations of the Magellanic Clouds have shown that cosmic ray
nucleons in the energy range below 100 GeV are almost certainly galactic
(\cite{sre93}), while the observations made with the OSSE and COMPTEL
instruments aboard CGRO have provided strong evidence that
cosmic ray electrons are galactic
(\cite{schlick97}, see also \cite{faz66}). Therefore the diffuse
Galactic \gr emission 
tells us about the propagation of cosmic rays from their sources to the 
interaction regions and thus complements the direct particle measurements
by balloon and satellite experiments.

The greater sensitivity and spatial and energy resolution of EGRET
compared with SAS-2 and COS-B permit a much more detailed
analysis of the diffuse Galactic \gr emission than was possible
with the earlier experiments. The spatial and spectral
distribution of the diffuse emission within $10^\circ$ of the Galactic plane
have recently been compared with a model calculation of this emission
which is based on realistic interstellar
matter and photon distributions and dynamical balance (\cite{hu97}),
i.e. cosmic rays having the same spectrum and composition everywhere
in the Galaxy and having an intensity which follows the surface density of
thermal gas convolved with a Gaussian with dispersion $\sigma =1.7\,{\rm kpc}
\ -\ 2.0\,{\rm kpc}$ (\cite{ber93}). The distribution of the
total intensity above 100 MeV agrees surprisingly well with the model 
predictions. However, at higher energies above 1 GeV the model systematically
underpredicts the \gr intensity. If the model is scaled up by a factor 1.6, 
the model prediction and the observed intensity above 1 GeV agree well.
Thus the
model displays a deficit of $\sim$38\% of the total observed emission
which depends, if at all, only weakly on location. At energies above 1 GeV
around 90\% of the model intensity is due to $\pi^0$-decay, i.e. hadronic
processes, and only 10\% is due to interactions of electrons.

There are a number of possible explanations for this deficit:

-- A miscalibration of EGRET could cause an overestimation of the
intensity above 1 GeV. This possibility is 
highly unlikely. Point sources generally show power-law spectra
without spectral hardening above 1 GeV.
It would require an extreme level
of cosmic conspiracy for a calibration error to mimic a general power-law
behavior in the spectra of cosmic \gr sources.

-- The kinematics of $\pi^0$ production may be poorly understood.
Detailed Monte-Carlo calculations have shown
(\cite{mori97}), that models based on the current knowledge
of particle interactions do not give very different results for the $\pi^0$
spectra than do simple isobar plus scaling descriptions (\cite{der86}).
It is unlikely that the cosmic ray nucleon spectrum in the solar vicinity
is softer
than that elsewhere in the Galaxy. The local cosmic ray spectrum samples
sources within a few kpc in distance and a few times 10$^7$ years in time.
Since the observed deficit
appears to be independent of Galactic longitude, the sources
of cosmic rays within a few kpc from the sun would have to be different from
those in the inner Galaxy and those in the outer Galaxy.
We have also tested and verified that the uncertainties in the local
interstellar cosmic ray spectrum below a few GeV are by far not sufficient to
account for the deficit. The uncertainties arising from our limited knowledge
of the cosmic ray nucleon spectrum and the nucleon-nucleon interaction
kinematics can be estimated to be on the order of a few percent.

-- There may be unresolved point sources which contribute strongly
at higher \gr energies. The only known class of objects with appropriate
spectra is pulsars. Based on the properties of the six identified
\gr pulsars it has been found that unresolved pulsars would indeed
contribute mainly between 1 GeV and 10 GeV (\cite{pohl97}). However,
to account for all
the deficit it is required that more than 30 pulsars be detectable
by EGRET as point sources. This can be compared with less than
10 unidentified \gr sources which are not variable (\cite{mcl96})
and which show pulsar-like spectra (\cite{merck96}). Also, the latitude
distribution of the \gr emission from unresolved pulsars is inconsistent
with that of the observed emission. Unresolved pulsars will contribute
6-10\% of the observed \gr intensity above 1 GeV and around 3\% in the
energy band between 100 MeV and 1 GeV (\cite{pohl97}), and thus they can account only for a small fraction of the high-energy \gr deficit. 

All in all the effects described above can account only for a small
fraction of the deficit, or can add only small systematic
uncertainties.
In this paper we will investigate whether the remaining deficit 
of 30-35\% may be caused by inverse Compton emission of high-energy electrons.
Leptonic processes contribute only around 10\% of the intensity
above 1 GeV in the model of Hunter et al. (1997), which corresponds
to $\sim $6\% of the total observed intensity. Thus the leptonic contribution 
would have to be increased to 35-40\% of the total observed emission to explain 
all the deficit.  

The Galactic distribution of cosmic ray electrons is intimately linked to that
of their sources. The recently-found evidence of X-ray synchrotron radiation
from the four supernova remnants SN1006 (\cite{koyama95}), RX J1713.7-3946
(\cite{koyama97}), IC443 (\cite{keohane97}), and Cas A (\cite{allen97})
supports the hypothesis that Galactic cosmic ray electrons are accelerated
predominantly in SNR. X-ray synchrotron radiation
implies TeV \gr emission from the comptonization of the microwave
background at a flux level which depends only on the average 
magnetic field strength (\cite{pohl96}), and indeed the detection of the
remnant SN1006 at TeV energies has been announced recently 
(\cite{tanimori97}).
Interestingly, there is no clear observational proof that the nuclear component
of cosmic rays is accelerated likewise in SNR. The acceleration of cosmic ray
nucleons in SNR should lead to observable flux levels at TeV
energies (\cite{drury94}), however with a spectrum different from that of the
leptonic emission. The generally tight upper limits for TeV emission from
the nearest SNR (\cite{lessard95,buckley}) are in conflict with simple
shock acceleration models for cosmic ray nucleons in SNR.

Most of the radio synchrotron spectra of SNR can be well represented by
power-laws with indices around $\alpha \simeq 0.5$, corresponding to 
electron injection indices of $s\simeq 2.0$ (\cite{gr95}). This is
in accord with predictions based on models of particle acceleration
(\cite{be87}). However, it is different from the electron injection 
spectral index of $s=2.4$ which has been inferred from the locally
observed electron spectrum (\cite{sk94}) and which subsequently has been used
in the model of Galactic \gr emission of Hunter et al. (1997).
The contribution of cosmic ray electrons to the Galactic \gr spectrum
at high energies depends strongly on their injection spectral index.
If the acceleration cut-off energy is high enough, the leptonic \gr emission
may even dominate at TeV-PeV energies (\cite{pp97}). A change in the
electron injection index by $\delta s =0.4$ could increase the inverse
Compton emissivities at a few GeV by an order of magnitude or more.

Let us suppose that for some reason
the local cosmic ray electron spectrum is different from the 
average electron spectrum in the Galaxy. Then the following scenario
appears viable:
the bulk of cosmic ray electrons is accelerated in SNR with an injection
index around $s\simeq 2.0$. The leptonic \gr emission at a few GeV
would be much stronger than in the Hunter et al. model, and it may 
explain a substantial fraction of the discrepancy between their model
and the observed spectra. So if we would find a mechanism or an effect
which would cause the local electron spectrum to be different from the 
Galactic average, we may in a second step reassess the \gr spectra produced
by Galactic cosmic rays without having to assume an electron injection
index of $s=2.4$.

It has been noted before that the spatial distribution of cosmic ray
sources affects the locally observable spectra (\cite{cl79,ls82a}).
As far as electron acceleration in SNR is concerned, there is no evidence that
the star formation activity and thus the SNR production rate
in the solar vicinity is significantly less than the Galactic average.
In the next section we will show that for cosmic ray electrons, unlike
nucleons, the local spectra above a certain energy can deviate
from the Galactic average, even if the spatial distribution of cosmic ray
accelerating SNR is homogeneous. This is a result of the discrete nature
of SNR both in space and in time.
We will use this finding in the third
section to model the high-energy \gr excess as result of inverse Compton
emission of cosmic ray electrons, albeit with a harder
injection spectrum than conventionally assumed.

\section{The time-dependence of the local electron spectrum}
The spectrum of cosmic ray electrons in terms of the contributions of discrete
sources like supernova remnants (SNR) has been discussed before (\cite{cl79}).  
These authors have investigated the case of continuously active sources and 
have concluded that one needs sources situated within a few hundred parsecs of
the solar system, in order that the energy losses of electrons do not induce
a cutoff in the energy spectrum. Since the required number of active sources 
exceeds that of supernova remnants by an order of magnitude, SNR were found
unlikely to be the only source of cosmic ray electrons between 1 GeV and 1 TeV.
In that paper the diffusion coefficient had been assumed independent of energy.
With the usual energy dependence $D \propto E^{(0.3-0.6)}$ some of the
statements in the Cowsik and Lee paper would have to be relaxed.

In this section we will also consider the finite lifetime of SNR or
other possible cosmic ray accelerators together with the random distribution
of SNR in space and time. The latter induces a time dependence in
the local electron spectrum at higher energies, which stems from the
Poisson fluctuations in the number of SNR within a certain distance and within a
certain time interval. As we will see, the discreteness of sources does
not simply cause a cutoff in the electron spectrum, but makes it
time variable and thus unpredictable beyond a certain energy.

Since effects of the discreteness of sources show up only at
higher particle energy, we may describe the propagation of cosmic ray electrons
at energies of 1 GeV to 1 TeV
by a simplified transport equation
\begin{equation}
{{\partial N}\over {\partial t}} - {{\partial}\over {\partial E}} (bE^2\, N)
-D\,E^a\, \nabla^2 N = Q
\end{equation}
where we consider continous energy losses by synchrotron radiation and inverse 
Compton scattering, an energy-dependent diffusion coefficient $D\,E^a$, and 
a source term $Q$.
The Green's function for this problem can be
found in the literature (\cite{gs64}).
\begin{equation}
G(r,r',t,t',E,E')={{\delta\left( t-t' +{{E-E'}\over {b\, E\, E'}} \right)}
\over {bE^2\, \left( 4\pi\,\lambda(E,E')\right)^{3/2} }}\ 
\exp \left(-{{(r-r')^2}\over {4\, \lambda(E,E')}}
\right)
\end{equation}
where
\begin{equation}
\lambda(E,E')= {{D}\over {b\,(1-a)}}\ \left(E^{a-1} -E'^{a-1}\right)
\end{equation}
In case of discrete sources the injection term $Q$ is a sum over all 
sources. For an individual source showing up at time $t_0$ and injecting for a
time period $\tau$ we can write
\begin{equation}
Q_i=q_0\, E'^{-s}\, \delta(r')\, \Theta (t'-t_0)\, \Theta (t_0 +\tau -t')
\end{equation}
Without loss of generality we can set $t=0$ and obtain
\begin{equation}
N=q_0\, E^{-s} \int_{-{{1}\over {bE}}}^0\ dt'\ 
{{\Theta (t'-t_0)\, \Theta (t_0 +\tau -t')\, \exp\left(-{{r^2}\over
{4\Lambda}} \right)}\over
{\left(4\pi \Lambda\right)^{3/2}\ ( 1+bEt')^{2-s}}}
\end{equation}
where
\begin{equation}
\Lambda= {{D\,E^{a-1}}\over {b\,(1-a)}}\ \left(
1-(1+bEt')^{1-a}\right)
\end{equation}
and $r$ is the distance between source and observer. $N$ is the contribution
to the local electron spectrum provided by a single source (SNR) at
distance $r$ which is (was) injecting electrons for a time period $\tau$ 
starting at $t_0$.
The local spectrum of electrons can now be obtained by summing the
contributions from
all individual sources. In our case 
the distribution of SNR is, for ease of exposition
and computation, assumed to be a
homogeneous disk of radius $r_s=15\,$kpc and half-thickness $z_s=0.14\,$kpc.
Other choices for the spatial distribution of SNR do not impose serious
changes in the results, as long as the distribution is not 
structured on sub-kpc scales.

The numerical procedure is as follows:
for each volume element $\delta V$ the expected number of SNR injecting
cosmic rays
within the time interval $\delta t$ is
\begin{equation}
N_{\lambda} = {{\delta V\ \delta t}\over {V_{tot}\ t_{inj}}}
\end{equation}
where $V_{tot}$ is the total volume of the source distribution and
$t_{inj}=55\,$years
is the inverse of the supernova rate (\cite{ca97}). $\delta t$
has to be the maximum lookback time in our model which is the
sum of the SNR lifetime $\tau$ and the electron energy loss time
$t_l ={{1}\over {b\,(1\,{\rm GeV})}}$ at 1 GeV, the lowest
energy considered here. At a given time
the actual number of SNR in that volume element is a Poissonian
random number with mean $N_{\lambda}$, and each of the SNR has a
birth date which is a random number uniformly distributed
within $[-\tau -t_l,0]$. The final electron spectrum is then derived by summing
over the contributions of the individual SNR per volume element and summing
over all relevant volume elements. We calculate 400 such `random' spectra and 
thus derive the distribution of possible spectra and their spread.

\placefigure{acct3}

In Fig.1 we show the resulting range of local electron spectra
compared with the observed spectra
(\cite{fer96,gol94,gol84,ta84,tai93}). The energy density of the ambient
photon fields plus that of the perpendicular component of the magnetic field
strength is in total taken to be 3.5 eV/cm$^3$. The changes in the Compton
cross section in the Klein-Nishima regime of optical and near-infrared
photon fields are neglected.
The diffusion coefficient is
$D=4\cdot 10^{27}\ {\rm cm^2\, sec^{-1}}$ at 1 GeV and increases with
energy to the power a=0.6, in accord with results for a two-dimensial diffusion
model fit to the local spectra of 13 primary and secondary cosmic ray
nuclei at rigidities between 1 GV and 10$^3$ GV (\cite{wlg92}). At these
energies electrons and nuclei will scatter off the same turbulence,
except for the helicity, and thus their diffusion behaviour may be expected
to be similar. The injection spectral index of electrons is s=2.0, which
corresponds to the mean synchrotron spectral index of individual SNR 
(\cite{gr95}). 

The solar modulation of electrons at a few GeV energy has a strong effect
on the observed spectrum. Note the difference between the spectrum observed 
when the modulation level was high (\cite{gol94}) and the spectrum
observed during the passage of Ulysses over the solar south pole (\cite{fer96}).
We have crudely approximated the effect of solar modulation using
the force-field approach (\cite{ga68}) with $\Phi=400\,MV$. Thus we probably 
overestimate the modulation in case of the Ulysses data and underestimate
in case of the Golden et al. data. 

The spectra at a given time are not necessarily smooth. There is no
preference for the usual broken power-laws or power-laws
with exponential cut-offs. In fact the individual spectra are bumpy
above $\sim$ 50 GeV and some display step-like features. As an example
we show a particular spectrum as dash-dotted line in Fig.1, which is 
slightly on the low side between 10 GeV and 100 GeV, where it suffers further
softening before it abruptly hardens at 300 GeV. 

Below 10 GeV the local electron spectrum is well
determined. Between 10 GeV and 100 GeV it varies with time by a factor of 
2 or 3, and above 100 GeV is completely unpredictable. Changes in the 
absolute numbers for the diffusion coefficient and the radiative energy losses
do not change the basic behaviour, but can shift the transition between
weak and strong variability to lower or higher
energies. If the energy dependence of the diffusion coefficient is weaker,
i.e. $a< 0.6$, the transition between weak and strong variability will be 
faster, and vice versa (a slower transition for higher powers than a=0.6). 
For comparison we have indicated the result for an energy dependence
of the diffusion coefficient, $D\propto E^{0.33}$. 

As shown in Fig.1, the high-energy data for the local electron flux are in accord with an injection spectral index of s=2.0, though in a model with
steady injection and a smooth source distribution
these data would require an injection index of around 
2.4 (\cite{sk94}). Concerning the distribution of high-energy electrons,
the Galaxy would look like swiss cheese, with holes and regions of higher density. In the line-of-sight integrals, which are relevant for
comparison with the EGRET \gr data, the averaging over holes and
high-density regions will give the same result as a model with steady
injection, however with source index of 2.0 instead of 2.4.
At higher latitudes the line-of-sight will be so short that
regions of low or high electron density will be resolved. The leptonic \gr
spectra in direction of the Galactic poles should be relatively soft since 
the line-of-sight integral of the \gr emissivity will be dominated by the
soft local spectrum.

The absolute electron flux is reproduced if each SNR provides an energy
input of $10^{48}\,{\rm ergs}$ in the form of electrons, which is
one thousandth of the canonical value of $10^{51}\,{\rm ergs}$ for the kinetic
energy input per supernova. Taken over a lifetime of $10^5\,{\rm years}$ the
corresponding power of $5\cdot 10^{35}\, {\rm ergs/sec}$ is less than the
X-ray luminosity of SN1006 alone.

The time variability of the high energy cosmic ray spectrum will not be
related or even be synchronous to variability in the flux
of low-energy cosmic ray nucleons, which can be traced by cosmogenic
instable isotopes in sediments (\cite{so87,mh95,ko96}) or meteorites
(\cite{bo96}).

A few notes should be added. We have taken supernova explosions
to be completely independent of each other. One may expect some level of
correlation in OB associations and SNOBs, which would make 
the basic effect of time dependence even more dramatic, since the OB
associations and SNOBs would act as single sources with longer lifetime,
but much smaller frequency of occurrence. 

Another important point is that we have assumed 
that all electron sources produce the same spectrum.
In reality this doesn't need to be the case. Some SNR will produce
electrons with harder spectrum, and another group of SNR will provide
softer spectra. It may be that
the spectral form depends on the age of an SNR.
In fact the radio data show that SNR do have different synchrotron spectra
(\cite{gr95}). If we take the electron injection spectral index of an
individual SNR not as a fixed number, but a random variable following some 
probability function, the time-averaged spectrum (the dotted line in 
Fig.\ref{acct3}) will get a positive curvature. The level of time variability
on the other hand will get larger. The shaded regions in Fig.1, in which
the spectrum is contained for 68\%, respectively 
95\%, of the time, extend beyond those for fixed injection index.

A final note concerns secondary positrons and electrons. These particles
are generated subsequent to interactions of cosmic ray nucleons with ambient
gas, and hence the effect discussed does not apply and the local spectrum of secondary electrons will not vary. Thus the observed positron fraction will
also exhibit variability anticorrelated with that of the primary
electron spectrum.  
If we are indeed living in a hole in the distribution of high-energy
electrons then the positron fraction above, say, 20 GeV will be
above the level expected in steady injection models, if the gas
density within $\sim\,$1 kpc from the sun is not also sub-average.
This may explain the observed positron fraction in that energy range,
which is indeed slightly above the model predictions
(\cite{barwick97}).

We have seen that the discreteness of sources of cosmic ray electrons
causes a strong variability of the local electron spectrum at higher
energies. Therefore the high energy electron spectrum does not prescribe
our choice of electron injection spectrum in propagation models.

If we consider \gr emission in the Galactic plane, the line-of-sight
integral of the emissivity will correspond to an averaging over the different variability states, and hence the \gr intensity calculated with time-dependent
models will not differ significantly from the results of steady-state models.
The latter are much easier to compute and thus preferrable, but they will
in general not be able to reproduce the local high energy electron spectrum
correctly. On the other hand, 
in the absence of reliable data on the position and age of all nearby SNR
it is also impossible to calculate the time-dependent local electron spectrum
precisely, we can only infer the level of variability. Therefore we
feel that, if the acceleration of electrons occurs predominantly in SNR,
the \gr emission in the Galactic plane can be sufficiently well described
with steady-state models, provided the model is
{\bf not required to fit
the local electron data above $\sim\,$30 GeV}.
In the next section we will discuss a steady-state model for
the diffuse leptonic \gr emission for an injection index of 2.0.

\section{The propagation of electrons in the steady state case}
There is a richness of
literature on the topic of electron propagation in the Galaxy,
including analytical solutions for the one- or two-dimensional
diffusion and diffusion-convection problem (\cite{gs64,bs69,ls81,ls82,ps91}).
These solutions can be well described in their basic behavior by the
concept of the catchment sphere (\cite{we70}). The energy losses prevent
electron propagation farther from their source than a critical
distance $\rho$
which is defined by equality of the time scales for transport and for energy
losses, so that the spatial dependence of the Green's function of the problem
is basically a function which is a constant for distances less than $\rho$
and which is zero beyond $\rho$. If the transport is governed by diffusion
and if the energy loss terms do not strongly depend on location, 
then $\rho$ will not strongly depend on direction, and thus
a source of cosmic rays at the position $(x',y',z')$ would fill a sphere of
radius $\rho$ with cosmic rays. Hence we may separate the spatial
problem and the energy problem, and approximate the solution to the
spatial problem by a Gaussian function for the catchment sphere.
The Gaussian function is exact at higher energies where radiative energy
losses dominate, but is a crude approximation at very
low energies, where ionization and Coulomb losses are important.

We will include escape as a catastrophic loss term, which limits $\rho$
to some maximum value. This approximates the effect of a sudden increase
of the diffusion coefficient at a certain height above the Galactic
plane (\cite{ls82}). We regard this a better description than a finite boundary
with density and density gradient set to zero at $L_{halo}$, a few kpc
above the
Galactic plane, since for a physical escape solution the density
outside the diffusion region relates to that inside as
\begin{equation}
n_{out} \simeq {{ n_{in}\, L_{halo}}\over {\tau_{esc}\, \beta\, c}}
\simeq 10^{-4}\ n_{in} \left( {{E}\over {\rm GeV}}\right)^{0.6}
\end{equation}
and thus it is definitely not zero.
With a diffusion coefficient $D=D_0\, E^a$,
$E$ being the kinetic energy, we have
\begin{equation}
\rho (E) =\sqrt{4\, D_0\, E^a}\, \sqrt{\tau_{eff}}
\end{equation}
where
\begin{equation}
\tau_{eff}^{-1}=\tau_{esc}^{-1} + \tau_{loss}^{-1}
\end{equation}
We can write the differential number density of cosmic rays 
at position $(x,y,z)$ coming from a source at $(x',y',z')$ as
\begin{equation}
\delta N ={1\over {(\sqrt{\pi}\, \rho)^3\,|{\dot E}|}}\ 
$$ $$ \qquad \quad \times\int_E du\ Q(u)\,
{\rm exp}\left(-\smallint_E^u {{dv}\over {\tau_{esc}(v)\,| \dot E(v)|}}
\right)\ {\rm exp}\left(
-{{r^2}\over {\rho^2}}\right)
\end{equation}
where
\begin{equation}
r^2=(x-x')^2+(y-y')^2+(z-z')^2\ ,
\end{equation}
$\dot E$ is the energy loss term and $Q(E)$ is the source spectrum at position
$(x',y',z')$. Physically $\delta N$ is a propagator and it can be treated
like a Green's function.
Given the spatial distribution of sources $q(x',y',z')$
we obtain the cosmic ray spectrum at any position $(x,y,z)$ as
\begin{equation}
N(E)=\int dx'dy'dz'\ \delta N\, q(x',y',z')
\end{equation}
Our method thus enables us to calculate the three-dimensional distribution
of electrons resulting from an arbitrary three-dimensional distribution
of sources.

The individual SNR may accelerate electrons with slightly different spectra.
This would result in a positive curvature of the composite injection
spectrum (\cite{bb72}).
To demonstrate the effect of a possible dispersion of
the injection spectral index in the electron sources, we assume
that the injection indices for individual SNR follow a normal
distribution
\begin{equation}
P(s)={{1}\over {\sqrt{2\pi} \mu_s}} \exp \left(- {{(s-s_0)^2}\over
{2\, \mu_s^2}}\right)
\end{equation}
at the energy $E_0$. Radio spectral index measurements at a few GHz, 
corresponding to $E_0 \simeq 5\,$GeV, indicate $\mu_s \la 0.2$ (\cite{gr95}).
Then the source spectrum of
primary electrons is
\begin{equation}
Q_e=q_e\, \left(m_e c^2\right)^{s-1}\, E^{-s}\, \left({{E}\over {E_0}}
\right)^{0.5\,\mu_s^2\,\ln {{E}\over {E_0}}}
\end{equation}
For $\mu_s =0$ this reduces to the conventionally assumed single power-law.

The energy losses due to ionization and Coulomb interactions, bremsstrahlung
and adiabatic cooling, and synchrotron and inverse Compton emission
are well described by
\begin{equation}
-{\dot E}=7.2\cdot 10^{-13}\, n_H \tau\left[1+{\eta\over \tau}
{E\over {714\,m_e c^2}}+ {{\epsilon U_{mag}}\over {n_H \tau}}
\left({E\over {3727\,m_e c^2}}\right)^2\right]
\end{equation}
where we have used the following abbreviations
\begin{equation}
\epsilon= 0.75 \ + \ {{U_{rad}}\over {U_{mag}}}
$$ $$\eta = 1\,+\, n_H^{-1} \left( {{{\rm div}\, {\bf v}}\over {\rm 3\cdot 
10^{-15}\ sec^{-1}}}\right)\,+\,0.95 {{n_e}\over {n_H}}
$$ $$\tau = 1\,+\, 1.54 {{n_e}\over {n_H}}
\end{equation}
$n_H$ is the neutral gas density,
$n_e$ the density of ionized gas, ${\bf v}$ the bulk velocity of
electrons, and $U$ the energy density of the magnetic field and the ambient 
radiation field in ${\rm eV\, cm^{-3}}$.

We assume the magnetic field strength to be constant over the total volume
of the Galaxy with $B=10\,{\rm \mu G}$. As we shall see later, this value
leads to synchrotron emission consistent with observations.
The interstellar radiation field
can be calculated from the respective emissivities
for optical and infrared emission, and the microwave background emission
(\cite{ys91} and recent updates \cite{str}). The distribution of
ionized gas has been modelled on the basis of pulsar data (\cite{tc93}).
The derivation of the
distribution of neutral matter will be discussed below. For the propagation 
calculation all parameters of the interstellar medium are averaged over a
scale of 1 kpc to mimic the average environmental conditions of a cosmic
ray electron during its life time. A Galactic wind is assumed to operate
in the Galactic halo, such that adiabatic cooling outside the disk provides
similar energy losses as bremsstrahlung inside the disk, i.e. the energy loss
terms can be written independent of the spatial location within a
catchment sphere. Note that the energy loss terms for neighboring catchment 
spheres may be different, since they are averages over different volumes.
The radial extent of the Galaxy is taken to be 16.5 kpc. Note that the
computer time consumption scales as radial extent squared.
The calculation of the bremsstrahlung and inverse Compton emissivities is
described elsewhere (\cite{pohl94}).

The energy loss time scale, which determines
the radius of the catchment sphere $\rho$, can be understood as
an ensemble average
\begin{equation}
\bar{\tau} = {{\int_E^\infty dE'\ Q(E')
\ \int_{E'}^{E} du \ \dot u^{-1}}\over {\int_E^\infty dE'\ Q(E')}}
\end{equation}
where $Q(E)$ is the electron injection spectrum. This average age can be
approximated by 
\begin{equation}
 \tau_{loss}=  \int_{E_c}^E du \ \dot u^{-1}\ ,\ \ E_c=2.718\,E
\end{equation}
to better than $30\%$ accuracy except for the lowest energies.
For $E_c \le 714\,
\tau\,\eta^{-1}\,m_e c^2$, corresponding to $E\la 100\ {\rm MeV}$,
the average age $\bar{\tau}$ is larger than the e-folding
energy loss time scale. Here the energy losses will
also depend explicitely on $\rho$ since ioniziation and Coulomb interactions
occur only in the gas disk.
For ease of computation we will assume $\tau_{loss}$
to be constant
at these low energies, as if ionization and Coulomb 
interactions would not occur. This means we overestimate $\rho$ but
underestimate ${\dot E}$. The two effects work in opposite directions
but may not balance each other. We want to keep in mind that we are
using a crude approximation at low electron energies, which for this
paper however will have an impact only on the bremsstrahlung spectra 
at $\la 50\ {\rm MeV}$.

We neglect secondary electrons in our model. The locally observed
fraction of secondary electrons is of order 10\% (\cite{barwick97}), but
the fraction may be strongly dependent
on position and on the propagation behavior
of particles (\cite{schlick82}). There are basically two arguments
which may allow us to neglect the secondaries. Firstly, because
of the energy dependence of the cosmic ray secondary-to-primary ratios,
secondary electrons will have a production spectrum which is softer
than the production spectrum of cosmic ray nucleons by $\delta s \simeq 0.6$,
at least above 1 GeV electron energy. Thus secondary electrons will have a 
softer spectrum than primary electrons, so that their contribution to the
diffuse Galactic \gr emission can not lead to a hardening of the spectrum,
irrespective of the flux. The second argument concerns the luminosity.
Since the production cross sections for charged and for neutral pions are
of the same order, and the electrons take only about two thirds of the pion
energy, the source power supplied to secondary electrons is linked to 
the hadronic \gr luminosity. Only a fraction of the source power is channeled
back into \gr emission, since synchrotron radiation takes away some energy.
Due to the kinematical low-energy cut-off at $\sim$100 MeV in the secondary
production spectrum and the decrease of hadronic \gr luminosity with energy
above 1 GeV, the secondary contribution to leptonic \gr emission above
100 MeV will always
be limited to less than 10\% of the \gr luminosity due to $\pi^0$-decay
(\cite{pohl94}) and thus be negligible in this energy range. 

\subsection{The distribution of gas}
The three-dimensional distribution of thermal material in the Galaxy has been 
determined by deconvolution of H$I$ surveys for atomic gas and
CO surveys for molecular gas with the rotation curve of \cite{cl85}.
The H$I$ surveys include the Leiden-Greenbank
survey (\cite{bl83,bu85}), the Weaver-Williams survey (\cite{ww73}), the 
Maryland-Parkes survey (\cite{ke86}), the high-latitude Parkes survey
(\cite{chh}), and the Heiles-Habing survey (\cite{hh}).
The CO data are taken from the Columbia survey (\cite{da87}) as updated 
(\cite{dd97}). All these data are publicly available from either the ADC
or the CDS data bank. It should be noted that none of these surveys is
stray-light corrected. The level of uncertainty in the rotation curve,
the position of the sun, and non-circular motions of gas is high,
so that our deconvolution should be taken as a model rather than as a datum.

The general procedere in the deconvolution process is as follows: The rotation
curve is used to calculate the relation between distance and
line-of-sight velocity, which is then transformed into a
probability distribution
for distance on the basis of the actual velocity resolution of the surveys
and a turbulent velocity dispersion of 10 km/sec for CO and
25 km/sec for H$I$. 
This approach tends to smear out the gas distribution along the
line-of-sight, but relaxes most of the forbidden velocity problem.
The near-far ambiguity towards the inner galaxy is resolved by
dividing the intensity according to the amplitudes of Gaussian
probability functions for the distribution of gas normal to the Galactic
plane. These Gaussian probability functions are the convolutions of
the local gas distribution functions and the spatial resolution
function of the particular survey. The effective scale heights of gas
on the far side of the Galaxy are thus systematically larger than the
local values. As a result the gas tends to be more evenly distributed
over the Galaxy than in other derivations (\cite{hu97}) and the deconvolved
gas distribution on the far side of the Galaxy will be slightly smeared out
normal to the Galactic plane. This has no impact on the calculation of
the \gr emission since the column density of gas is always preserved, 
and it has also no impact on the cosmic ray propagation since our 
algorithm is based on the gas surface density, which is also preserved.

The H$I$ data are scaled under the assumption of a constant spin temperature
of $T_s =125\ $K. The obvious absorption features in the direction of
the Galactic Center have been replaced by a linear interpolation between
the neighboring velocity bins. The distribution of H$I$ normal to the
Galactic plane is assumed to be a Gaussian of dispersion
$z_c =0.12+0.023 (r-9.5)\,\Theta (r-9.5)$ kpc, where $r$ is the
galactocentric radius and $\Theta$ is a Heavyside function, with an
offset according to the warping of the H$I$ disk (\cite{bu76}). 

The CO-data are scaled with an X-factor of 1.25, which is the mean of
the best fit values in published papers on EGRET data analysis
(\cite{hu94,digel95,dig96,sm96,hu97}). In the inner kpc of the Galaxy the X-factor is reduced
to 25\% of its nominal value to account for the higher excitation temperature
and different metallicity (\cite{sod95,ari96}). The noise in the CO spectra is
preserved in the deconvolution process to keep the line-of-sight integral
of the density unchanged. The vertical CO distribution is assumed to be
a Gaussian of dispersion $z_c =0.074+0.03 (r-9.5)\,\Theta (r-9.5)$ kpc,
where $r$ is the galactocentric radius and $\Theta$ is a Heavyside function
(\cite{da87}). The position of the sun is assumed to be located
8.5 kpc from the Galactic Center and 15 pc above the plane (\cite{ha95}).

For a region of $\sim\, 20^\circ$ towards the Galactic Center and towards
the anticenter the kinematical resolution is insufficient, and the data have
been edited by hand. The distribution of gas in these two regions is basically 
an interpolation between the results for the adjacent regions. In the
anticenter region any excess over this interpolation has been evenly
distributed over all distances, while in the Galactic Center region any
excess is attributed to the Galactic Center. The distribution of gas in
the Galactic plane is shown in Fig.\ref{gas}.

\placefigure{gas}

\subsection{The spatial distribution of sources}

The true distribution of SNR in the Galaxy is not well known
as a result of selection effects and the absence of a proper
distance measure to the remnants. In many papers
(\cite{sj77,du88,bloemen93}) the cosmic ray distribution in the Galaxy 
has been estimated on the basis of a functional form for the SNR distribution
which fits the data for 116 remnants (\cite{ko74}). One of the general findings
in these studies is that the overall cosmic ray distribution is too steep
to explain the gradual slope of the \gr emissivity over the Galactic radius
(\cite{str88,sm96}).

Here we use a revised functional form for the SNR distribution (\cite{cb96}),
which fits the data for 194 remnants
\begin{equation}
f(r)=\left({{r}\over {r_{\odot}}}\right)^{1.69\pm 0.22} 
\exp\left(-(3.33\pm 0.37) {{r-r_{\odot}} \over { r_{\odot}}}\right)
\end{equation}
where $r_{\odot}=$8.5 kpc denotes the distance between the sun and
the Galactic Center. The vertical distribution of SNR is taken to be
box-shaped with a half thickness of 150 pc.

Our model allows us to use true 3D source distributions. We know that the
Galaxy has structure in the form of spiral arms, a bulge and so forth.
We have thus folded a spiral arm model (\cite{gg76,vallee}) with the radial
SNR distribution as described above to investigate the influence of spiral arm
structure. In the Georgelin and Georgelin model the Galaxy has four symmetric
arms with pitch angle of around 12$^\circ$. We have rescaled their model to 
$r_{\odot}=$8.5 kpc. Each spiral arm is described by a Gaussian of dispersion
500 pc, i.e. a FWHM of 1177 pc. The spiral arm model is
normalized in azimuth, so that the integral $\smallint_0^{2\pi} r\,d\phi$
of that model yields unity, and then folded with the radial SNR 
distribution. The normalization is required to preserve the radial distribution
of SNR. A face-on view of the resultant source distribution is shown
in Fig.\ref{src}.

\placefigure{src}

\section{Results}

In this section we show results for two choices of the spatial distribution of sources, the pure SNR distribution and the SNR distribution
folded with a model of the spiral arms in the Galaxy. We will also show
results for two choices of injection index, at first a fixed index s=2.0 for all
sources, and then a normal distribution of indices with mean 2.0 and
dispersion 0.2. We will not vary any other parameter in the propagation model,
and stick to the best fit values given by Webber, Lee and Gupta (1992).
In a forthcoming paper we will extend our model to nucleons and determine
the propagation parameters self-consistently which fit the \gr data and
the local spectra of primary and secondary cosmic rays.
In this paper our emphasis is to show that an injection index of around s=2.0
for cosmic ray electrons is sufficient to explain the observed spectrum
of diffuse high-energy \grs, while leaving all other parameters unchanged.

\subsection{The local electron spectra}

We have shown in Sec.2 that if electrons are accelerated in SNR, their local 
spectra above $\sim$ 30 GeV will strongly depend on time, so that the direct
electron measurements above this energy may deviate from the average 
electron spectrum. As a result, we are not required to choose the electron 
injection index according to the directly observed electron spectrum
above $\sim$ 30 GeV. Below a few GeV, on the other hand, the locally
observed spectra are strongly affected by solar modulation,
for which we do not have reliable models, so that 
only over roughly one decade in energy does the local electron spectrum
provide clear data. The data from direct electron measurements
compared with the steady-state spectra of our model are shown in 
Fig.\ref{local_both}. The model
spectra fit the data reasonably well in the relevant energy range up
to 30 GeV. 

\placefigure{local_both}

\subsection{The \gr emission}

As an example we show in Fig.\ref{hunt_snr} the spectra of the bremsstrahlung
and inverse Compton emission in the direction of the inner Galaxy
for the case of sources with injection indices following a normal distribution
of mean s=2.0 and dispersion $\mu$=0.2.
The figure includes the results of our model,
the observed spectrum, and a template of the $\pi^0$-decay spectrum
(\cite{der86}).
At around 5 GeV
the intensity due to leptonic processes is higher than that
due to hadronic $\pi^0$ decay. Our model assumes that the power-law behaviour
of the electron injection spectra persists to $\sim$20 TeV. 
The high-energy \gr spectrum will rather directly reflect structure in the 
electron source spectra.
If for example the true source spectrum would deviate from a simple
power-law above 1 TeV, the Inverse
Compton spectrum would show corresponding features above 50 GeV.

\placefigure{hunt_snr}

When we consider the latitude distribution of the \gr intensity above
1 GeV, which is done in Fig.\ref{plotb_both}, we find a high level
of agreement. The fraction of the total diffuse intensity, which is due to 
leptonic emission, is almost constant between b=0$^\circ$
and b=10$^\circ$. It
goes up from $\sim$6\% in the Hunter et al. model,
$\sim$10\% in our model for an injection index of s=2.4, to 30\% - 48\%
for an injection index of s=2.0,
enough to explain all the \gr excess.

\placefigure{plotb_both}

We can compare the longitude distribution of the observed diffuse
\gr emission above 1 GeV
to that of the leptonic contribution in our model. This is done
in Fig.\ref{plotl_both}. It is obvious that the 
galactocentric gradient of the leptonic \gr emission in our model is stronger
than in the data. This is the case for all cosmic ray propagation models
which are based on the SNR distribution (\cite{wlg92,bloemen93}).
While towards the inner Galaxy the
fraction of the total diffuse intensity, which is due to leptonic emission,
is around 35\% - 52\%,
far enough to explain all the \gr excess, towards the outer Galaxy the
leptonic contribution in our model accounts only for roughly
two thirds of the excess.

\placefigure{plotl_both}

The SNR distribution of Case and Bhattacharya has a zero at 
galactocentric radius r=0 which causes
the double peak structure in the model intensity towards the inner Galaxy. 
Such double peak structure is not visible in the data, so that this effect
may result more from the specific choice of mathematical function in the fit
of the SNR distribution than astrophysical reality. We can
generally say that a flatter SNR distribution would beget a
better harmony of the longitude
distribution of observed emission and model. Interestingly,
one study of the SNR distribution, which was
not based on a specific form of the radial profile, indicates very long
radial scale lengths up to 9 kpc (\cite{li91}). 

We have therefore
investigated the impact of the fit uncertainties in the SNR distribution. Within 1 $\sigma$ in both parameters the SNR distribution may be 
\begin{equation}
f(r)=\left({{r}\over {r_{\odot}}}\right)^{1.91} 
\exp\left(-2.96 {{r-r_{\odot}} \over { r_{\odot}}}\right)
\end{equation}
Here we show the results for the spiral arm model only. 
For similar local electron spectra the leptonic
contribution to the diffuse Galactic \gr emission above 1 GeV can be 33\% -47\%
while the center/anticenter contrast in the model agrees with the observed
one to better than 25\%. As can be seen
in Fig.\ref{plotb_flat}, the latitude distribution of the leptonic
contribution varies little. The longitude distribution of the leptonic 
contribution in Fig.\ref{plotl_flat} shows that the overall gradient
is reasonably well reproduced, but the double hump structure towards the
inner Galaxy remains as well as a general overprediction 
around $\vert l\vert \approx 45^\circ$.
This double hump structure however is a consequence of the mathematical
function chosen to fit the SNR distribution and not a consequence of the
SNR distribution itself.

\placefigure{plotb_flat}

\placefigure{plotl_flat}

\subsection{The synchrotron emission}
A further constraint on our model is the synchrotron flux
towards the North Galactic Pole (NGP). Available data at 408 MHz
(\cite{has82}), at 820 MHz (\cite{berk}),  and at
1420 MHz (\cite{rr86}) can be corrected for zero level
uncertainties, the contributions of the microwave background
and unresolved extragalactic sources (\cite{rr88}), and contributions from
the Coma cluster (\cite{sst86}). Here we do not use data at frequencies below 
100 MHz, which we expect to be affected by free-free absorption.
In Fig.\ref{synngp} we compare the synchrotron intensity
predicted by our model
with the data in direction of the NGP. It can be seen that for a
total magnetic field strength of 10$\mu$G there is good agreement.

\placefigure{synngp}

In our model the FWHM of the z-distribution of synchrotron emission
at 1420 MHz is $\sim$1.1 kpc, which is the value typically found
for edge-on galaxies (\cite{hummel}).

\section{Summary and discussion}

In this paper we have investigated whether cosmic ray electrons can be
responsible for the recently observed high intensity of diffuse Galactic
\gr emission above 1 GeV. Models based on the locally observed
cosmic ray spectra underpredict the observed intensity by nearly
40\% (\cite{hu97}). One feature of these models is the relatively soft
electron injection spectral index of s=2.4 (\cite{sk94}), which is required
to account for the local electron spectrum above 50 GeV.

The recent detection of non-thermal X-ray synchrotron radiation
from the four supernova remnants SN1006 (\cite{koyama95}), RX J1713.7-3946
(\cite{koyama97}), IC443 (\cite{keohane97}), and Cas A (\cite{allen97})
supports the hypothesis that Galactic cosmic ray electrons are accelerated
predominantly in SNR. We have shown in this paper that, if this is
indeed the case, the local electron spectra above 30 GeV are
variable on time scales of a few hundred thousand years. This variability 
stems from the Poisson fluctuations in the number of SNR in the solar
vicinity within a certain time period. While the electron spectra below 10
GeV are stable, the level of fluctuation increases with electron energy,
and above 100 GeV the local electron flux is more or less unpredictable.

With that time variability in mind we have seen that an electron injection
index of s=2.0 is consistent with the data of the direct particle 
measurements if SNR are the dominant source of cosmic ray electrons.
In fact, both the radio spectra of individual SNR (\cite{gr95}) and the hard spectrum of the inverse Compton emission at high latitudes (\cite{chen96})
would better harmonize with an injection index of s=2.0 instead of
s=2.4.

We have then presented a three-dimensional steady-state diffusion
model for cosmic
ray electrons, based on the propagation parameters which have
been derived from similar models for cosmic ray nucleons. 
While being entirely consistent with the local electron electron flux up to 
$\sim $ 30 GeV energy and with the radio synchrotron spectrum towards
the North Galactic Pole, the leptonic contribution to the diffuse Galactic
\gr emission above 1 GeV
in the Galactic plane increases from $\sim$6\% in the model
of Hunter et al., $\sim$, 10\% in our model for an injection index s=2.4,
to 30-48\% for an injection index s=2.0 depending on the assumed
spatial distribution
of SNR and depending on whether some dispersion of injection spectral 
indices is allowed. An electron injection index s=2.0 can therefore
explain the bulk of the observed \gr excess over the predictions of the
Hunter et al. model.

While the latitude distribution of the leptonic \gr emission is fully
consistent with that of the observed emission, we find that the longitude
distribution deviates from the observed one. In our model the contrast
between the Galactic Center and the anticenter is stronger than in the data.
A similar effect can be found in all cosmic ray propagation models
which are based on the SNR distribution (\cite{wlg92,bloemen93}). Note
that in the Hunter et al. model only the emissivity spectrum is taken
according to a propagation calculation (\cite{sk94}), while the spatial distribution of the emissivity is scaled according to the distribution of
thermal gas. 

We have seen that the fit uncertainties
in the SNR distribution of Case and Batthacharya allow us to use a
flatter profile, which leads to a better agreement between model
and data in the longitude distribution (see Fig.\ref{plotl_flat}).
Thus the gradient problem may be simply the result of an inappropriate
choice of radial SNR distribution or lack of error propagation, respectively.
This flatter profile would also harmonize better with the results
of Li et al. (1991), who found very large radial scale lengths of up to
9 kpc for the galactic distribution of SNR.
Other possible sources of systematic errors are discussed below.

This gradient problem is unlikely to be caused by additional 
thermal matter. Very cold (3$^\circ$K) molecular gas has recently been
discussed as candidate for baryonic dark matter (\cite{pfe94}). If organised
in small clumps (\cite{copfe94}) the probability of finding absorption
features in the spectra of bright background objects would be small and
the clumps would easily evade detection (for a review see \cite{copfe97}). 
However, the thermal gas mainly affects the bremsstrahlung and only indirectly
the inverse Compton emission, which dominates above 1 GeV \gr energy, and
thus has little influence on the gradient in the total leptonic \gr
emission above 1 GeV.

If our model for the interstellar radiation field were wrong, then it
would have only a limited influence on the gradient, since the Galaxy
acts as a fractional calorimeter at high electron energies (\cite{pohl94}),
which channels the electron source power directly into synchrotron and
inverse Compton emission with a ratio corresponding to that of the
energy densities in the magnetic field and the photon field.
Any radial variation of the magnetic field strength will be directly reflected
in the Center/anticenter contrast of the synchrotron intensity,
so that the radio surveys would constrain the parameter space here.

It is our personal view that the uncertainties in the radial 
distribution of SNR are large enough so that the overly strong gradient
in our model may simply be the result of an inappropriate choice of SNR 
distribution. The artificial double peak structure towards the inner Galaxy
is an example of systematic effects which arise from
possibly ill-defined fits to the SNR distribution.

Our findings indicate a potential problem in the determination of the
extragalactic \gr background (\cite{sre97}). The standard method uses
a linear regression analysis of observed intensities and model predictions
(with the model of Hunter et al.) to extrapolate to zero Galactic intensity.
If at higher \gr energies the inverse Compton emission is indeed much stronger 
than assumed by Hunter et al., then a large fraction of it will be attributed
to the extragalactic \gr emission. Our steady-state model predicts an
intensity of
high-latitude inverse Compton emission above 1 GeV at a level of $\sim 40\%$ of 
the extragalactic background intensity, while Hunter et al. assume much
smaller values. The intensity difference between the inner Galaxy and the
outer Galaxy at medium latitudes ($40^\circ - 50^\circ$) would be around
10\% of the extragalactic background intensity, depending on the choice
of model for the electron source distribution (SNR distribution).
On the other hand, if we are indeed living in a region of temporarily low
flux of high-energy electrons, we would also expect the intensity of inverse
Compton emission towards the Galactic Poles to be less that the steady-state
value, so that the true level of Galactic \gr intensity at high latitudes
is difficult to assess.

The spectrum of inverse Compton emission,
s$\simeq$1.85 in our model, is somewhat harder
than the s$\simeq$2.1 of the extragalactic background. It has been noted before
that the average \gr spectrum of the identified AGN is softer than that of
the background (\cite{pohl97b}), which indicates a problem with the idea
that the background is mainly due to unresolved AGN. Now, since the presently
determined background spectrum may be substantially contaminated with
hard inverse Compton emission, the true background spectrum may
be softer than s=2.1, which would probably relieve the spectral
discrepancy with the average spectrum of resolved AGN (\cite{pohl97b}).
Anyway, the systematic uncertainty in the spectral index of the
extragalactic \gr background emission is much higher than the
statistical uncertainty.

In a forthcoming paper we shall discuss the distribution of
synchrotron emission in more detail. We shall also use a truly
three-dimensional calculation of the interstellar photon fields based
on COBE/FIRAS data. Finally we shall discuss cosmic ray nucleons in parallel
with electrons to derive the propagation parameters self-consistently.

\acknowledgements

The EGRET Team gratefully acknowledges support from the following:
Bundesministerium f\"{u}r Bildung, Wissenschaft, Forschung und Technologie
(BMBF), Grant 50 QV 9095 (MPE); NASA Cooperative Agreement NCC 5-93 (HSC); 
NASA Cooperative Agreement NCC 5-95 (SU); NASA Contract NAS5-96051
(NGC); and NASA Contract NAS5-32490 (USRA).

\clearpage

\plotone{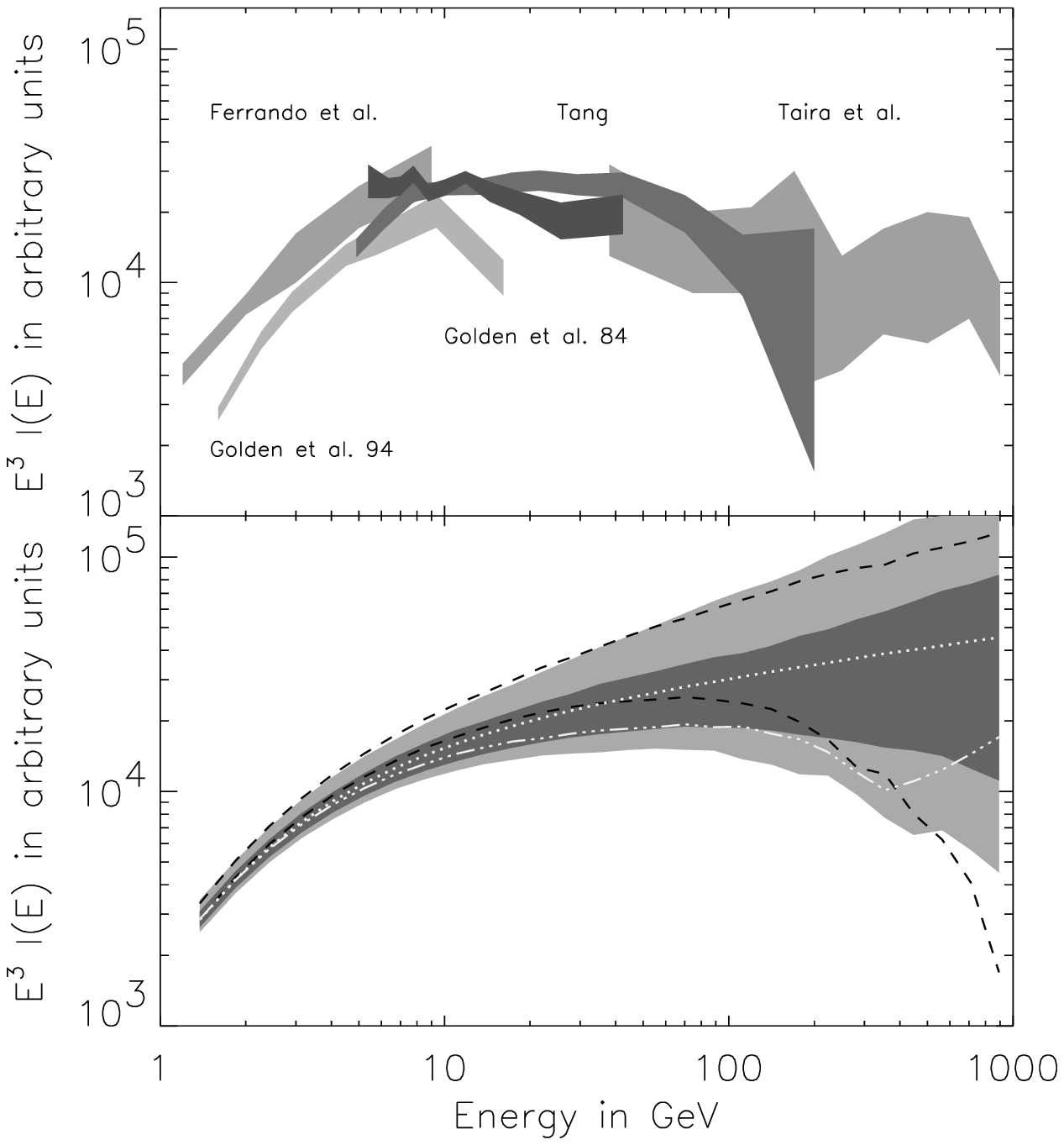}

\clearpage

\begin{figure}
\caption{The locally observed electron spectra in the upper panel 
compared with the range of possible spectra in our model.
The parameters of the model are given in the main text.
For each experiment the 1 $\sigma$ uncertainty range is indicated by a grey-shaded band which connects the data points at the mean energies of the
corresponding energy bins. The scatter between the results of different experiments indicates the level of systematic uncertainties. 
The range of possible spectra in our model is given by the grey-shaded bands
in the lower panel. During 68\% of the time the locally observed
spectra will be in the dark grey shaded region, and during 95\% of
the time they will be within the light grey shaded region. The dashed 
black curve
gives the 68\% range for a weaker energy dependence of the diffusion
coefficient (a=0.33 instead of a=0.6) to show the influence of this
parameter. The white dash-dotted line shows one of the 400 random spectra as
a particular example of what may be observed. 
The white dotted line indicates the time-averaged spectrum.
The effect of solar modulation is taken into account for all
model spectra using a force-field parameter $\Phi=400\ {\rm MV}$
(\cite{ga68}).
The data are not in conflict with the range of
possible spectra in our model.
\label{acct3}}
\end{figure}

\clearpage

\begin{figure}
\plotone{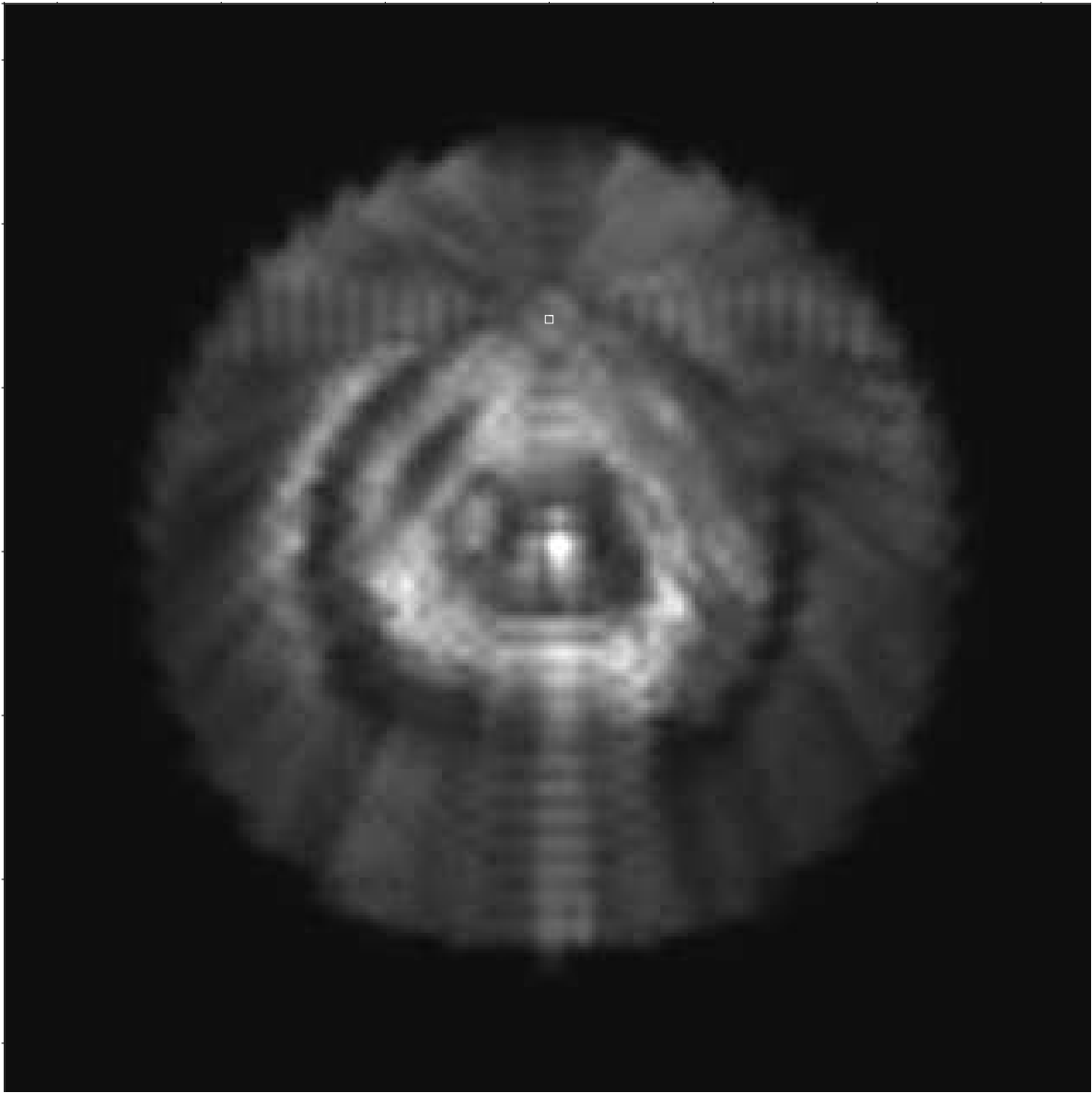}
\caption{Face-on view of the surface density of gas in the Galaxy.
The square indicates the position of the sun.
The color scale is linear between
surface mass densities of -2 $M_\odot \, pc^{-2}$ and
30 $M_\odot \, pc^{-2}$. The plot includes atomic, molecular, and ionized
gas integrated from -500 pc to 500 pc height above the plane. 
\label{gas}}
\end{figure}

\clearpage

\begin{figure}
\plotone{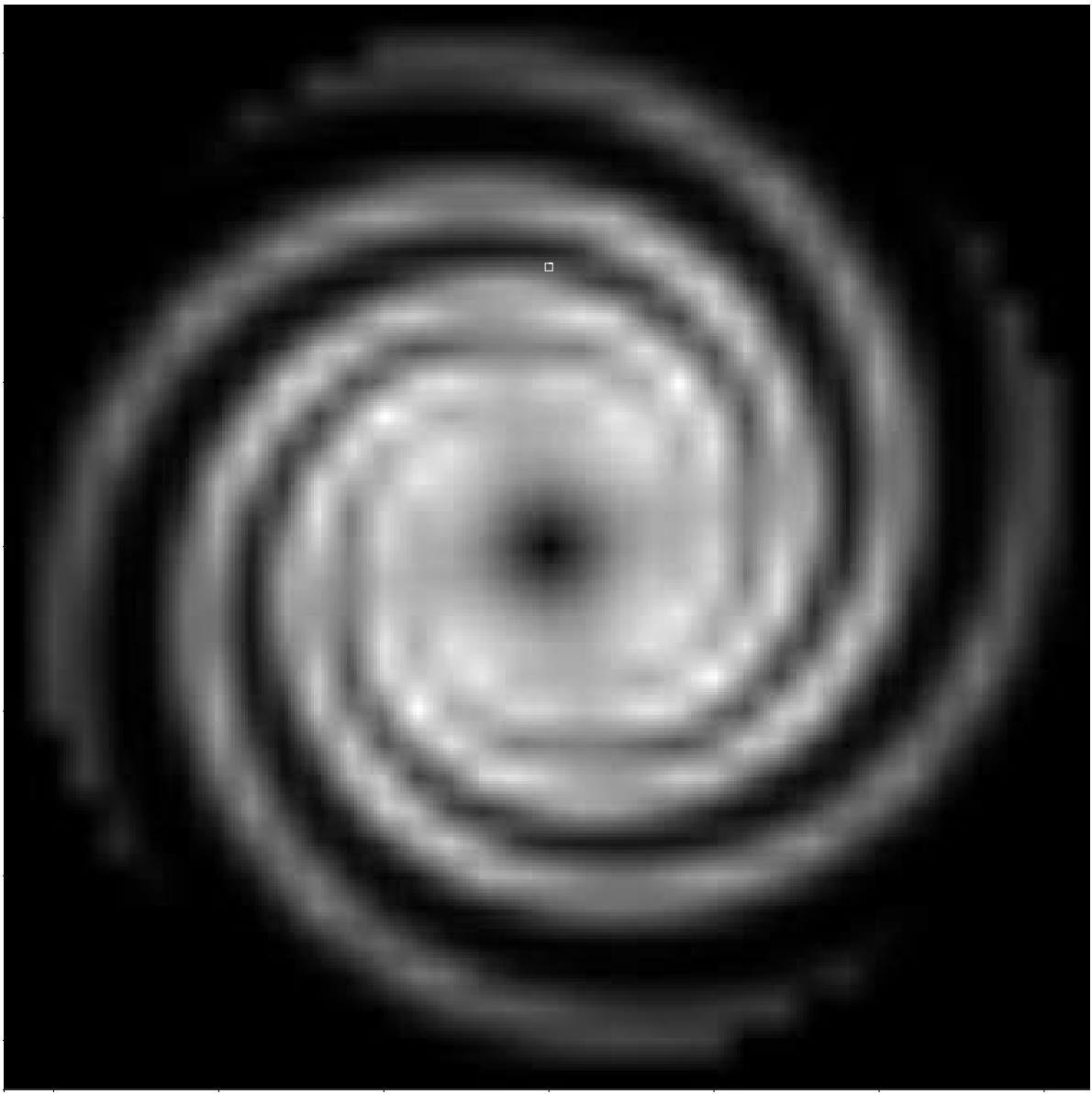}
\caption{Face-on view of the SNR distribution in the Galaxy
according to the spiral arm model.
The square indicates the position of the sun.
The color scale is linear between
zero and peak value. Note that the sun is located in an interarm region.
\label{src}}
\end{figure}

\clearpage

\begin{figure}
\plotone{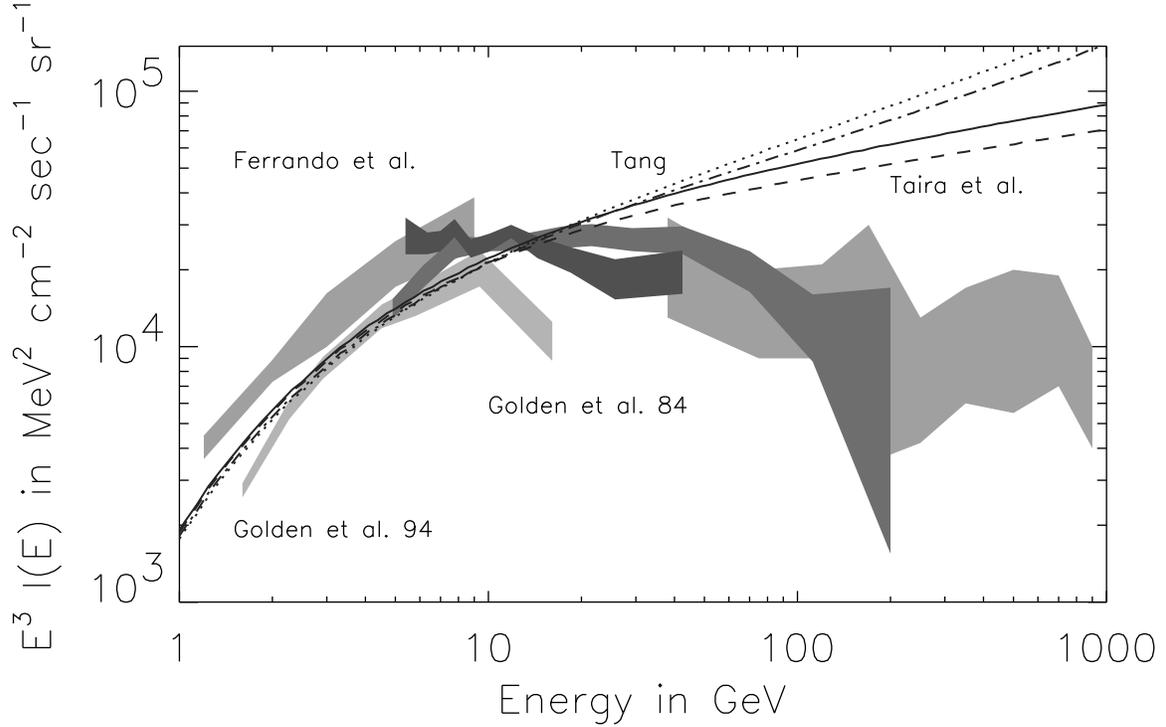}
\caption{The data for the local flux of cosmic ray electrons
compared with the steady state model spectrum.
Solar modulation is taken into account using a
force-field parameter $\Phi = 400\,{\rm MV}$ (\cite{ga68}). 
For the case of the pure SNR distribution as electron
source distribution, the solid line is for
sources with fixed injection index s=2.0, while the dotted
line is for source indices with a normal distribution of
mean 2.0 and dispersion 0.2. When the SNR distribution
in spiral arms is taken as source distribution, we obtain
the dashed line for
sources with fixed injection index s=2.0 and the dash-dotted
line for source indices with a normal distribution of
mean 2.0 and dispersion 0.2.\label{local_both}}
\end{figure}

\clearpage

\begin{figure}
\plotone{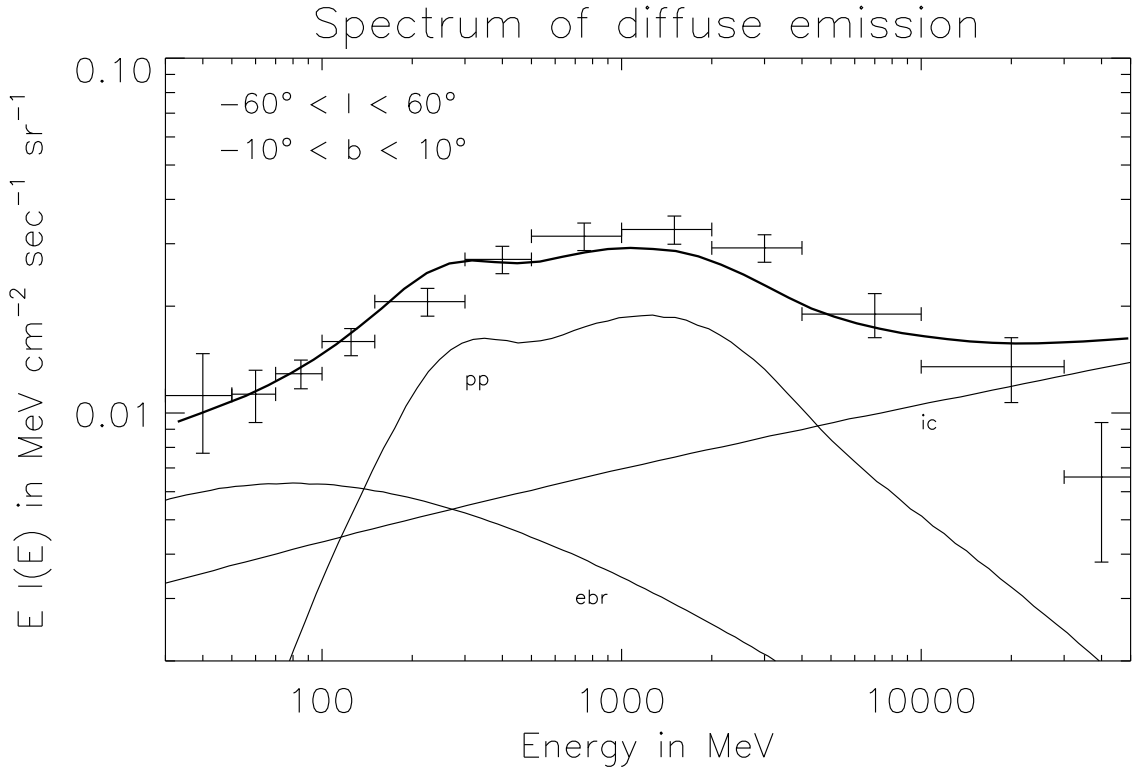}
\caption{The \gr intensity in the direction of the inner Galaxy.
The data points are taken from Hunter et al. (1997). The error bars
include an estimate for the systematic error of 8\%, which 
accounts for the uncertainty in the energy-dependent
correction of the spark chamber efficiency (\cite{esp98}).
The data are compared with bremsstrahlung (ebr) and Inverse Compton (ic)
spectra from our model, here on the basis of sources with
injection indices following a normal distribution of 
mean 2.0 and dispersion 0.2 and the spatial distribution of SNR in spiral arms.
The $\pi^0$-component is a template and not a model.
\label{hunt_snr}}
\end{figure}

\clearpage

\begin{figure}
\plotone{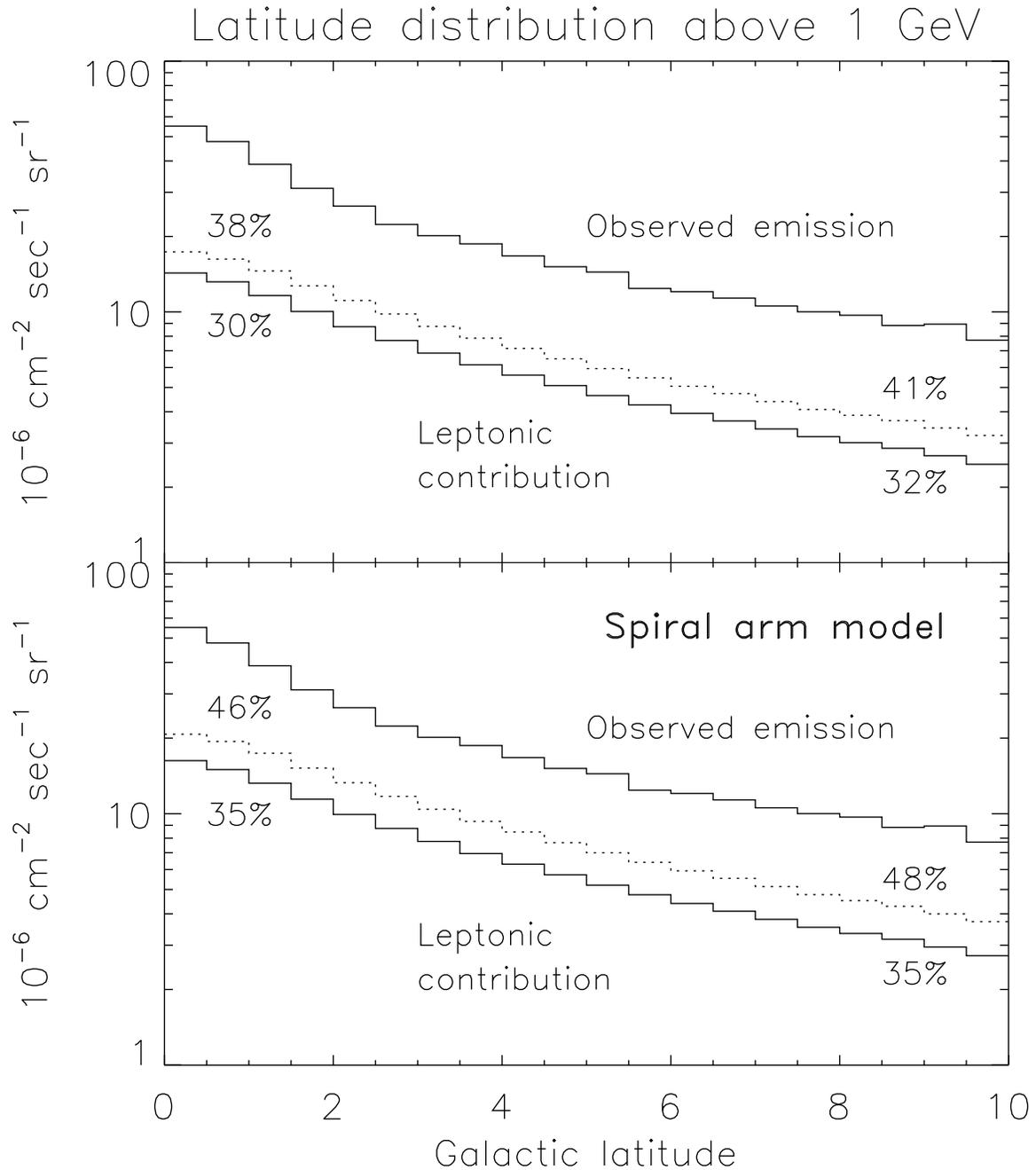}
\caption{The latitude distribution of the \gr emission above 1 GeV 
compared with the model prediction for the leptonic contribution.
Again the solid line is for
fixed injection index, and the dotted line is for a distribution of indices
with dispersion 0.2. The numbers give the percentage of the observed
emission in certain directions which is due to leptonic processes.
The top panel displays results for the pure
SNR distribution as electron source distribution, while the bottom panel 
is for a source distribution of SNR in spiral arms.
\label{plotb_both}}
\end{figure}

\clearpage

\begin{figure}
\plotone{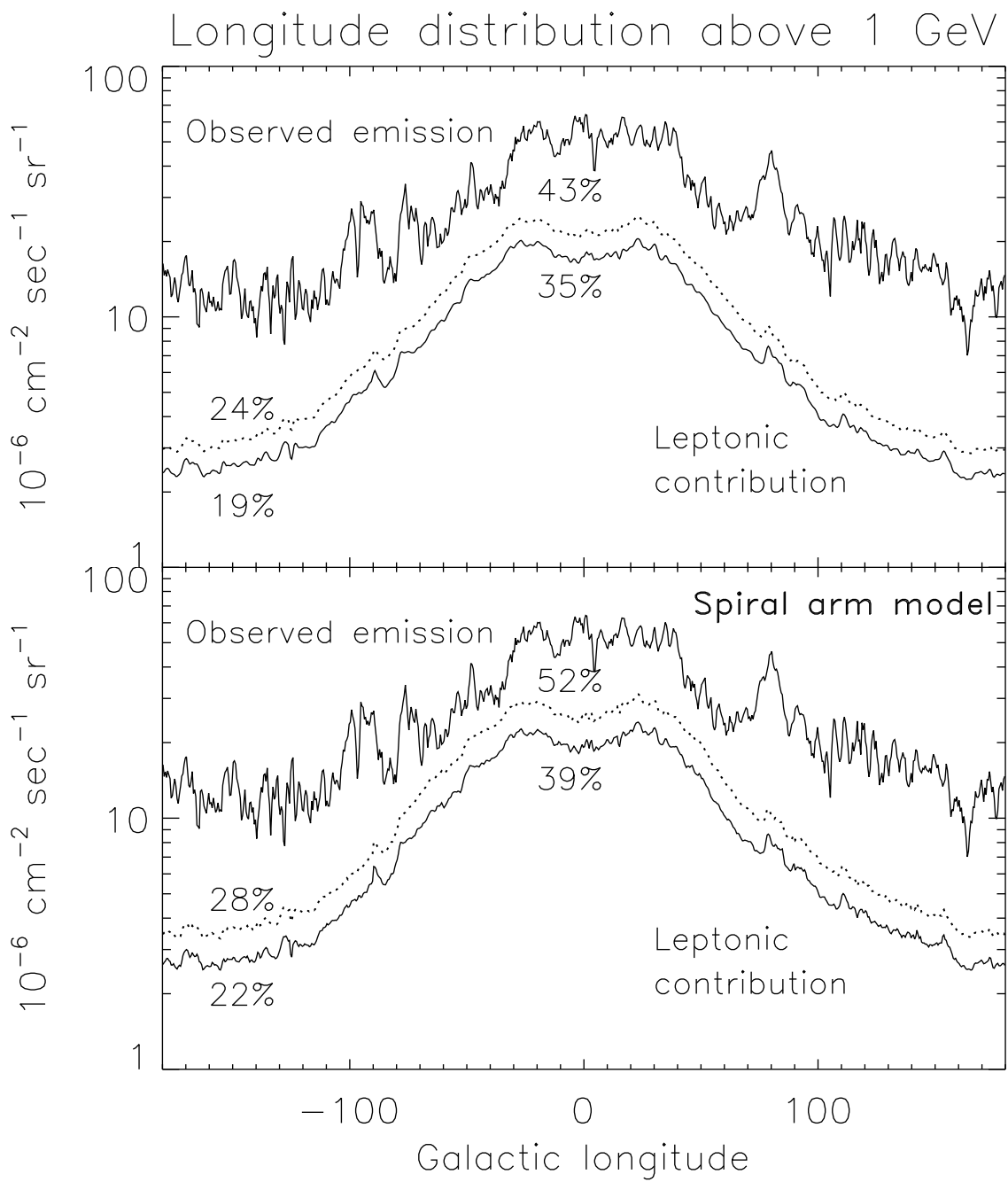}
\caption{The longitude distribution of the \gr emission above 1 GeV.
The upper solid line shows the observed distribution according to Hunter
et al. (1997). The distribution of the leptonic emission in our model
is shown for comparison, here for the SNR distribution as source  distribution.
Again the solid line is for fixed injection index, and the dotted line is for
a distribution of indices with dispersion 0.2. The numbers give the
percentage of the observed emission in certain directions which is due
to leptonic processes. The top panel is for the pure SNR distribution
and the bottom panel is for a source distribution of SNR in spiral arms.
\label{plotl_both}}
\end{figure}

\clearpage

\begin{figure}
\plotone{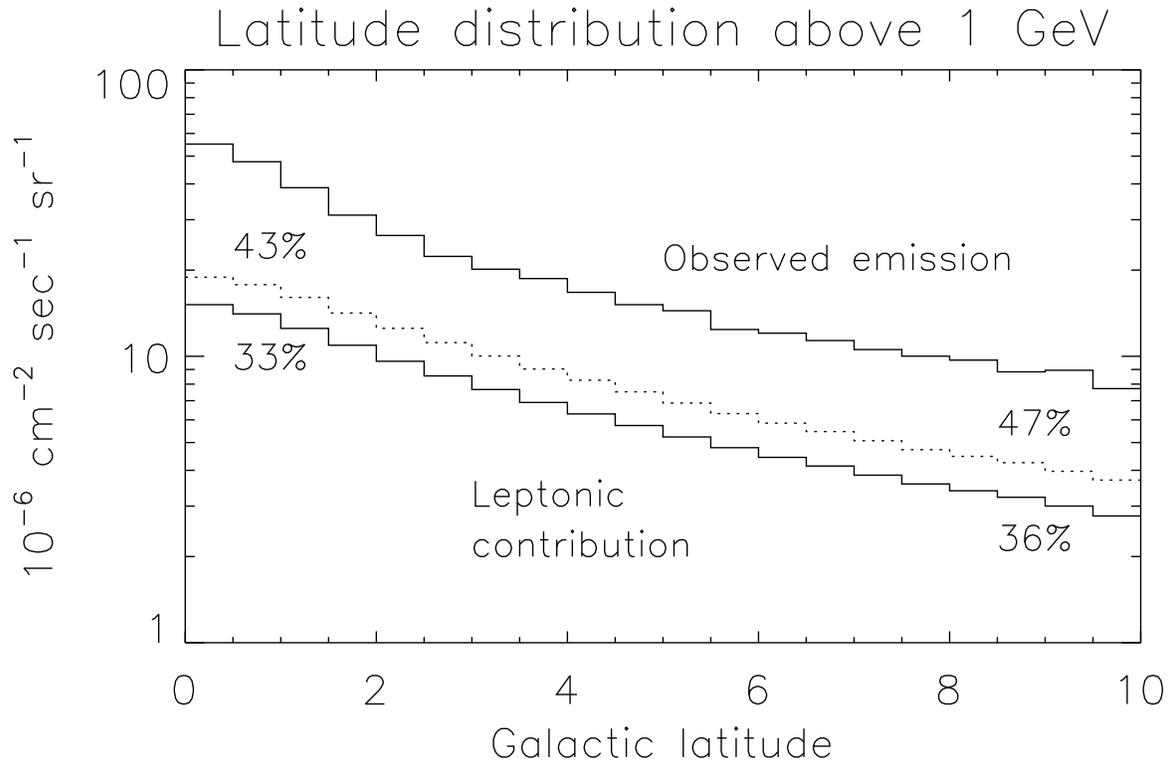}
\caption{The latitude distribution of the \gr emission above 1 GeV as in
Fig.\ref{plotb_both} except that here the flatter SNR distribution
in spiral arms has been used. The indicated percentage of leptonic
contribution of 33\% - 47\% is sufficient to account for all the excess.
\label{plotb_flat}}
\end{figure}

\clearpage

\begin{figure}
\plotone{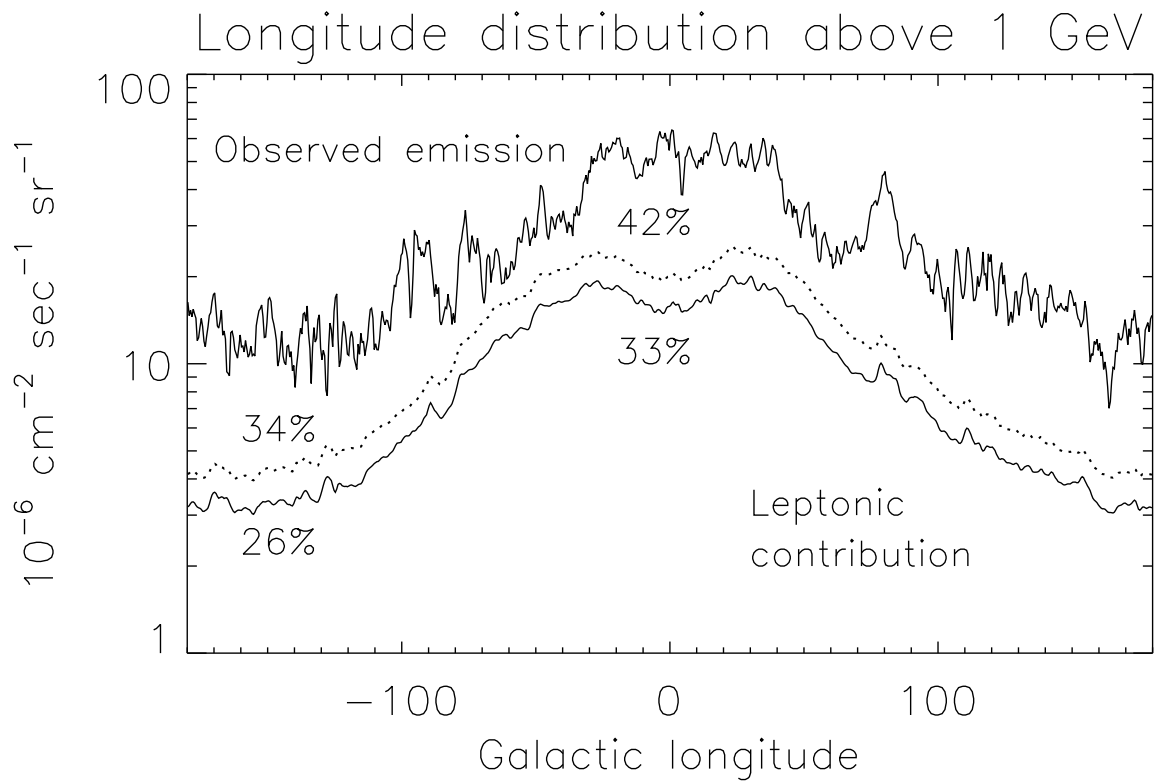}
\caption{The longitude distribution of the \gr emission above 1 GeV as in
Fig.\ref{plotl_both} except that here the flatter SNR distribution
in spiral arms has been used. The overall gradient is well reproduced.
The double hump structure towards the inner Galaxy and the overprediction 
around $\vert l\vert \approx 45^\circ$
indicate that the mathematical profile in the fit to the SNR 
distribution may be ill-defined.
\label{plotl_flat}}
\end{figure}

\clearpage

\begin{figure}
\plotone{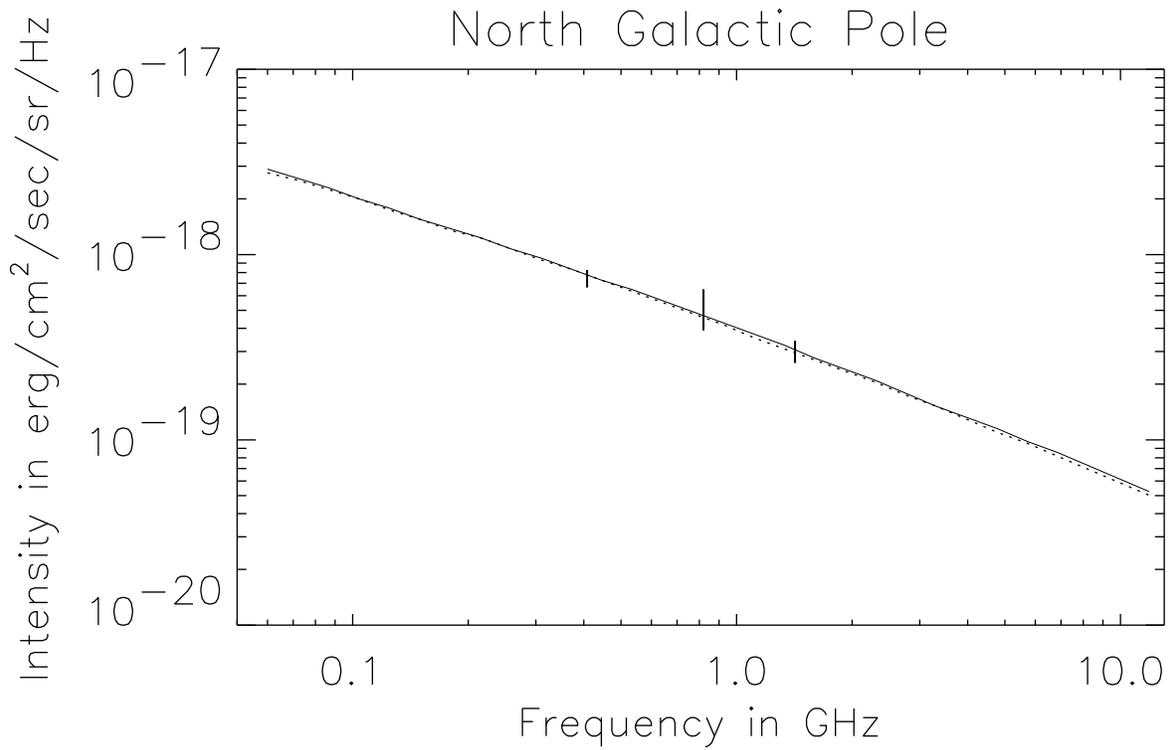}
\caption{The synchrotron intensity in the direction of the North Galactic Pole.
The error bars indicate data from surveys with reasonable zero level 
calibration. The solid line is the model spectrum based on the pure SNR
source distribution while the dotted line is for the SNR
distributed in spiral arms.
\label{synngp}}
\end{figure}

\end{document}